\documentclass[letter, 10pt, conference]{ieeeconf}      % Use this line for a4
\IEEEoverridecommandlockouts                              % This command is only
\usepackage{graphics} % for pdf, bitmapped graphics files
\usepackage{epsfig} % for postscript graphics files
\usepackage{amsmath} % assumes amsmath package installed
\usepackage{amssymb}  % assumes amsmath package installed

\usepackage{url}
\usepackage[ruled, vlined, linesnumbered]{algorithm2e}
\usepackage{verbatim} 
\usepackage{soul, color}
\usepackage{lmodern}
\usepackage{fancyhdr}
\usepackage[utf8]{inputenc}
\usepackage{fourier}
\usepackage{array}
\usepackage{makecell}
\usepackage{graphicx}
\usepackage{multirow}
\usepackage{adjustbox}
\usepackage{float}
\usepackage{hyperref}

\usepackage{graphicx}
\usepackage{subcaption}

\SetNlSty{large}{}{:}

\makeatletter

\newcommand{\Rom}[1]{\expandafter\@slowromancap\romannumeral #1@}
\makeatother

\title{\LARGE \bf
SpotKube: Cost-Optimal Microservices Deployment \\ with Cluster Autoscaling and Spot Pricing
}

\author{Dasith Edirisinghe, Kavinda Rajapakse, Pasindu Abeysinghe and Sunimal Rathnayake% <-this % stops a space 
\\ Department of Computer Science \& Engineering \\
University of Moratuwa, Sri Lanka \\
{\tt\small\{dasith.18, kavindar.18, pasindua.18, sunimal\}@cse.mrt.ac.lk} \\
}

\begin{document}

\maketitle
\thispagestyle{plain}
\pagestyle{plain}

%%%%%%%%%%%%%%%%%%%%%%%%%%%%%%%%%%%%%%%%%%%%%%%%%%%%%%%%%%%%%%%%%%%%%%%%%%%%%%%%
\begin{abstract}

Microservices architecture, known for its agility and efficiency, is an ideal framework for cloud-based software development and deployment, particularly when integrated with containerization and orchestration systems that streamline resource management. However, the cost of cloud computing remains a critical concern for many organizations, prompting the need for effective strategies to minimize expenses without compromising performance. While cloud platforms like AWS offer transient pricing options, such as Spot Pricing, to reduce operational costs, the unpredictable demand and abrupt termination of spot VMs introduce considerable challenges. By leveraging containerization alongside intelligent orchestration, it is possible to optimize microservices deployment costs while maintaining performance requirements.

We present SpotKube, an open-source, Kubernetes-based solution that employs a genetic algorithm for cost optimization. Designed to dynamically scale clusters for microservice applications on public clouds using spot pricing, SpotKube analyzes application characteristics to recommend optimal resource allocations, ensuring deployments remain cost-effective without sacrificing performance. Its elastic cluster autoscaler adapts to changing demands, gracefully managing node terminations to minimize disruptions in system availability. Evaluations conducted using real-world public cloud setups demonstrate SpotKube’s superior performance and cost efficiency compared to alternative optimization strategies.

\end{abstract}

{\small{\href{https://github.com/SpotKube/SpotKube}{\textit{GitHub: https://github.com/SpotKube/SpotKube}}}}


%%%%%%%%%%%%%%%%%%%%%%%%%%%%%%%%%%%%%%%%%%%%%%%%%%%%%%%%%%%%%%%%%%%%%%%%%%%%%%%%
% ------------------------- Introduction ------------------------------------------------
\section{INTRODUCTION}
Microservices architecture has gained widespread popularity for its agility, flexibility, and efficiency in cloud-based software development \cite{TheFutur71:online}. By breaking complex applications into manageable services, this approach allows for independent development, deployment, and scaling. 

However, the operational cost of running containerized microservices can be significant, and a cost-effective container orchestration solution, combined with transient pricing models like AWS Spot Pricing is essential.

Moreover, optimal resource utilization is critical to optimizing costs in microservices deployment. Yet, the dynamic nature of these applications, coupled with the unpredictable termination of spot instances, makes achieving this optimization challenging. Therefore, a thorough characterization of applications and careful resource analysis are essential.

Numerous studies have focused on solving distinct use cases such as microservice application characterization \cite{jindal2019performance}, cost optimization through transient pricing models for microservice applications \cite{gu2021scheduling}, and cost-optimized deployment strategies in public cloud environments \cite{Bento_Araujo_Barbosa_2023}. However, there remains untapped potential for a comprehensive framework that facilitates the deployment of microservice workloads in a public cloud environment, incorporating application characteristic analysis to deliver maximum monetary benefits to consumers via cost-optimized cluster autoscaling and transient pricing while simultaneously ensuring robust performance. \par

We propose SpotKube, an open-source Kubernetes-managed service that optimizes the deployment cost of microservices applications. SpotKube optimizes deployment costs through an application characterization tool that determines optimal configurations based on user-defined SLOs and genetic algorithm-based cost-optimized cluster autoscaling, leveraging transient pricing models like AWS Spot Pricing in public cloud environments.

In this paper, we evaluate SpotKube with representative microservices applications in a public cloud setup and compare its effectiveness with other optimization strategies. Our results show that SpotKube significantly reduces costs while meeting performance requirements.

Key Contributions:

\begin{enumerate}
    \item Measurement-driven reactive approach for cost-optimal microservices application deployment using spot instances
    \item An Elastic Cluster Autoscaler, powered by an optimization engine based on a genetic algorithm, aims to find the optimal balance between cost and performance.
    \item SpotKube, an open-source Kubernetes managed service for public clouds, integrates all of the above components
\end{enumerate}
The paper is organized as follows: Section II reviews related work. Section III details the SpotKube architecture and implementation. Sections IV describe the experimental setup and results. Section V discusses key findings, and Section VI concludes the paper and outlines future work.

% ------------------------- Related Works ------------------------------------------------
\section{RELATED WORKS}

This section presents a comprehensive literature review on the following aspects including microservice application characterization, cost modeling, auto-scaling, and container scheduling.

\subsection{Application Characterization}

Application characterization involves analyzing the performance of an application and identifying its optimal configurations to ensure optimal resource utilization and application performance. \par

Jindal et al. \cite{jindal2019performance} introduced a novel tool, Terminus, which employs a methodology encompassing a series of load tests to assess resource usage and calculate Microservice Capacity (MSC), defined as the maximum rate of HTTP or gRPC requests that can be served while adhering to Service Level Objectives (SLOs). \par

Han et al. \cite{han2020refining} proposed a framework for microservices placement in cloud-native computing environments that utilizes empirical profiling to identify and respond to workload characteristics. The framework involves conducting profiling experiments on selected workloads to derive resource requirements and then performing microservices placement using a greedy-based heuristic algorithm.

\subsection{Cost modeling}

Cost modeling is required to estimate the deployment cost and identify an optimal combination of resources to deploy the microservices. \par

Jorn Altmann et al. \cite{altmann2014cost} comprehensively analyzed the cost factors in cloud computing of federated hybrid clouds and developed a cost model consisting of a fixed cost formula, a variable cost formula, and a total cost formula. 

A. Baldominos Gómez et al. \cite{baldominos2022aws} employed machine learning techniques to predict AWS EC2 spot instance prices using historical datasets and EC2 API data. The results indicated significant variability in forecasting performance across different instance types.  \par

\subsection{Auto-scaling and scheduling containers} 

Kubernetes is an open-source system for orchestrating and managing the lifecycle of containerized applications \cite{Kubernet50:online} and recent studies have proposed extended components for Kubernetes to support auto-scaling and cost-optimized scheduling of containers. \cite{sami2020fscaler, rodriguez2020container, yan2021hansel} \par

Sami et al. \cite{sami2020fscaler} proposed an agent named FScaler, which utilizes reinforcement learning to scale the container's instances horizontally and schedule the placement of newly created instances based on defined cost functions after studying the change in resource availability. \par

Rajkumar Buyya et al. \cite{rodriguez2020container} developed custom extensions for Kubernetes, including a scheduler, rescheduler, and autoscaler. The scheduler optimally places tasks on cluster resources with the required CPU and memory. The autoscaler adjusts the number of VMs based on workload, scaling out to meet demand and scaling in to reduce costs by utilizing underutilized VMs \par

M. Yan et al. \cite{yan2021hansel} developed a tool to improve the horizontal auto-scaling policy of Kubernetes using the SARSA-based multi-agent parallel training module. This module predicts the task arrival rate and guarantees the assurance of SLA (Service Level Agreement). The complexity and resource usage of the task scheduling algorithm is comparatively high, which is an overhead to the overall performance. \par

Brandon et al. \cite{thurgood2019cloud} introduces a Free and Open-Source Software (FOSS) solution based on Kubernetes (K8s) for autoscaling K8s worker nodes within a cluster to support dynamic workloads. The solution is based on VMware vSphere technology in a private cloud model, but it can be adapted to leverage FOSS platforms and scale into hybrid cloud architectures and edge computing. 

Shridhar G. Domannal et al. \cite{domanal2018efficient} propose an efficient cost-optimized scheduling algorithm for a Bag of Tasks (BoT) using the Modified Particle Swarm Optimization technique. The authors present a Neural network-based backpropagation algorithm to predict spot instance values to reduce the cost further. 

Gu, Haihua, et al. \cite{gu2021scheduling} propose DSMT, a dynamic task scheduling algorithm to minimize VM renting costs for microservice workflow applications. A critical component, the Spot Instance Monitor, handles abrupt terminations by reallocating tasks from failed spot instances to on-demand instances, balancing rental costs and sub-deadlines of interrupted tasks.

Bento, Araujo, Barbose, et al. \cite{Bento_Araujo_Barbosa_2023} have proposed CAAS a method that optimizes cost savings while maintaining availability in a microservices architecture. CAAS uses a multi-objective optimization strategy to balance cost and availability by scaling pods. CAAS achieves significant cost reductions compared to traditional pod autoscaling methods like Kubernetes Horizontal Pod Autoscaler (HPA)

% ------------------------- Architecture ------------------------------------------------
\section{METHODOLOGY}

\subsection{\textbf{Overview}} The primary objective of the proposed system is to determine the most cost-efficient and application-specific allocation of resources to a Kubernetes cluster while ensuring adherence to user-defined Service Level Objectives (SLOs) and maintaining optimal application performance.

\begin{figure*}[!h] 
    \centering
    \includegraphics[width=0.9\textwidth]{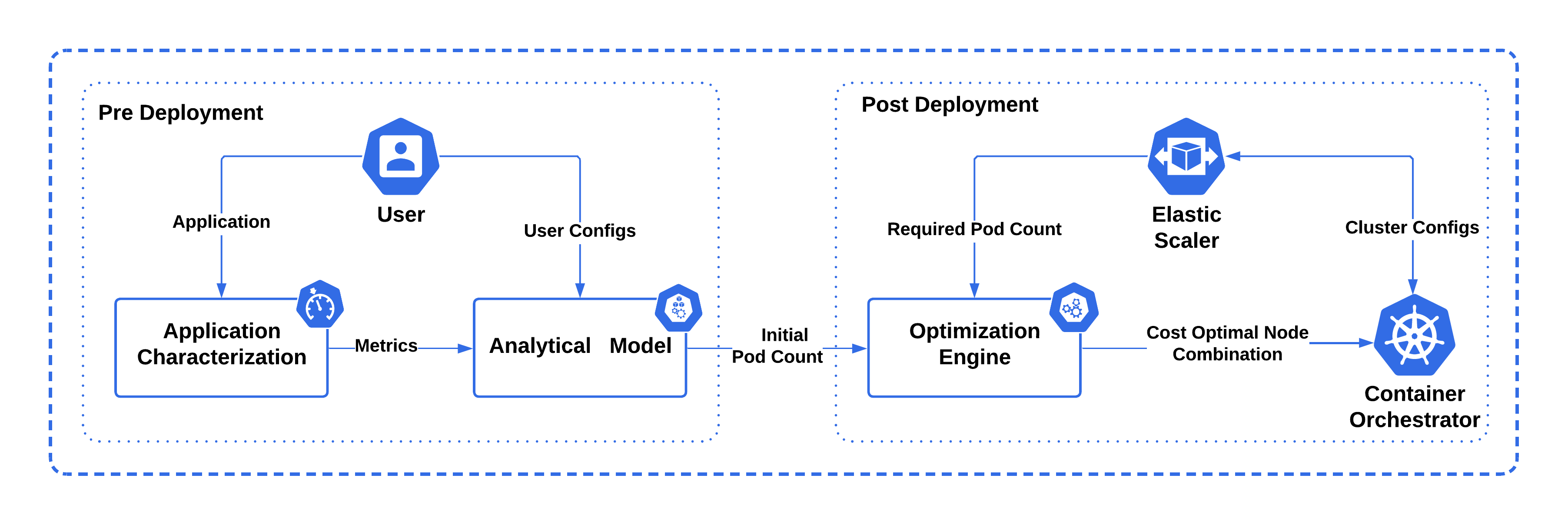} 
    \caption{High Level Architecture}
    \label{fig: High Level Architecture}
\end{figure*}

The architecture, as depicted in Figure \ref{fig: High Level Architecture}, summarizes the overall flow of our proposed approach and it can be divided into the following key sections.

\textbf{Pre Deployment}
\begin{itemize}
  \item \textbf{Microservice Application Characterization:} This phase involves analyzing the microservice application through a load test and collecting application-specific metrics, such as CPU usage and memory usage per given Requests Per Second (RPS).
  \item \textbf{Calculation of Initial Resource Requirements:} This phase entails mapping the application-specific CPU and memory requirements to the user-defined SLOs and calculating the initial number of pods required for deployment. This is done by the Analytical Model.
  \item \textbf{Cost optimal Resource Allocation:} This phase involves mapping the required pod count to available spot instances, calculating the cost-effective spot instance combination, and attaching it to the cluster. This process will be managed by the Optimization Engine.
\end{itemize}
Currently, the cluster is prepared for initial deployment. Following the successful deployment of the application, scaling the cluster becomes essential to ensure consistent performance and efficient cost management. The next phase of our approach is specifically designed to address these requirements. \\
\textbf{Post Deployment}
\begin{itemize}
 \item \textbf{Dynamic Scaling of the Cluster:} This phase is iterative and it involves dynamically attaching or detaching cluster resources based on incoming traffic, in a cost-effective manner while handling the abrupt termination of spot instances, using the Elastic Scaler and Optimization Engine.
\end{itemize}

% SpotKube can be extended to other cloud vendors and multi-cloud environments, but for demonstration purposes, we have integrated it with AWS services.

The following sections provide detailed explanations of each step, including the assumptions made, the solution, and the implementation process.

Assumptions and Constraints:

\begin{itemize}
  \item Only CPU-intensive applications are considered
  \item Only homogeneous pods are considered for deployment 
  \item Only one Service Level Objective(SLO) is considered.
  \begin{itemize}
        \item Minimum number of request count that should be handled by a microservice
    \end{itemize}
\end{itemize}

% \subsection{\textbf{Provisioning the Public Cloud Infrastructure}} One of the important tasks in the SpotKube pipeline is provisioning the public cloud infrastructures based on the user-provided configuration. To facilitate this process, the Provisioner has been designed to automate the provisioning workflow. \medskip

% \noindent\textbf{Provisioner:} Provisioning is a critical component of deploying microservice-based applications on the public cloud, as it involves building the necessary environment in public cloud environments for deploying the microservice application and running SpotKube services.  Provisioner automates this process, ensuring that the user's configurations are implemented consistently across all public cloud environments. The provisioner is responsible for creating the necessary infrastructure, including Virtual Private Clouds (VPCs), security groups, network interfaces, and management nodes. 

\subsection{\textbf{Microservice Application Characterization}} The process of application characterization represents a significant challenge due to the dynamic nature of each microservices application. In addition, it is essential to note that the approach adopted by SpotKube is limited to CPU-intensive applications. 

The application characterization tool comprised multiple steps. Firstly, the microservice-based application needs to be subjected to load testing using Locust\cite{LocustAm36:online}. Following this, CPU metrics are collected using Prometheus \cite{Promethe23:online}. \par

To automate the load-testing workflow, SpotKube has provided a script that takes care of the entire process. Initially, the user deploys the microservices application using Minikube and the provided configuration, followed by running the aforementioned script. The duration of the load test is determined by the application, with the test running until the failure rate surpasses a specific threshold. Finally, the tool outputs the load test results which includes CPU usage per given RPS.

It is important to mention that deploying pods with the same resource limits on different instance types results in almost the same performance \cite{jindal2019performance}. Hence regardless of the underlying infrastructure, load test results can be used for further analysis.

\subsection{\textbf{Calculation of Initial Resource Requirements}} By leveraging metrics collected through the application characterization tool the analytical model calculates the application-specific resource requirement and then determines the required number of pods that need to be deployed to the cluster, based on the Service Level Objectives (SLOs) specified by the user. Note that this pod count is the minimal number of pods that need to be initially deployed without violating the user-defined SLOs.

\begin{figure}[!h]
    \centering
    \includegraphics[scale=0.43]{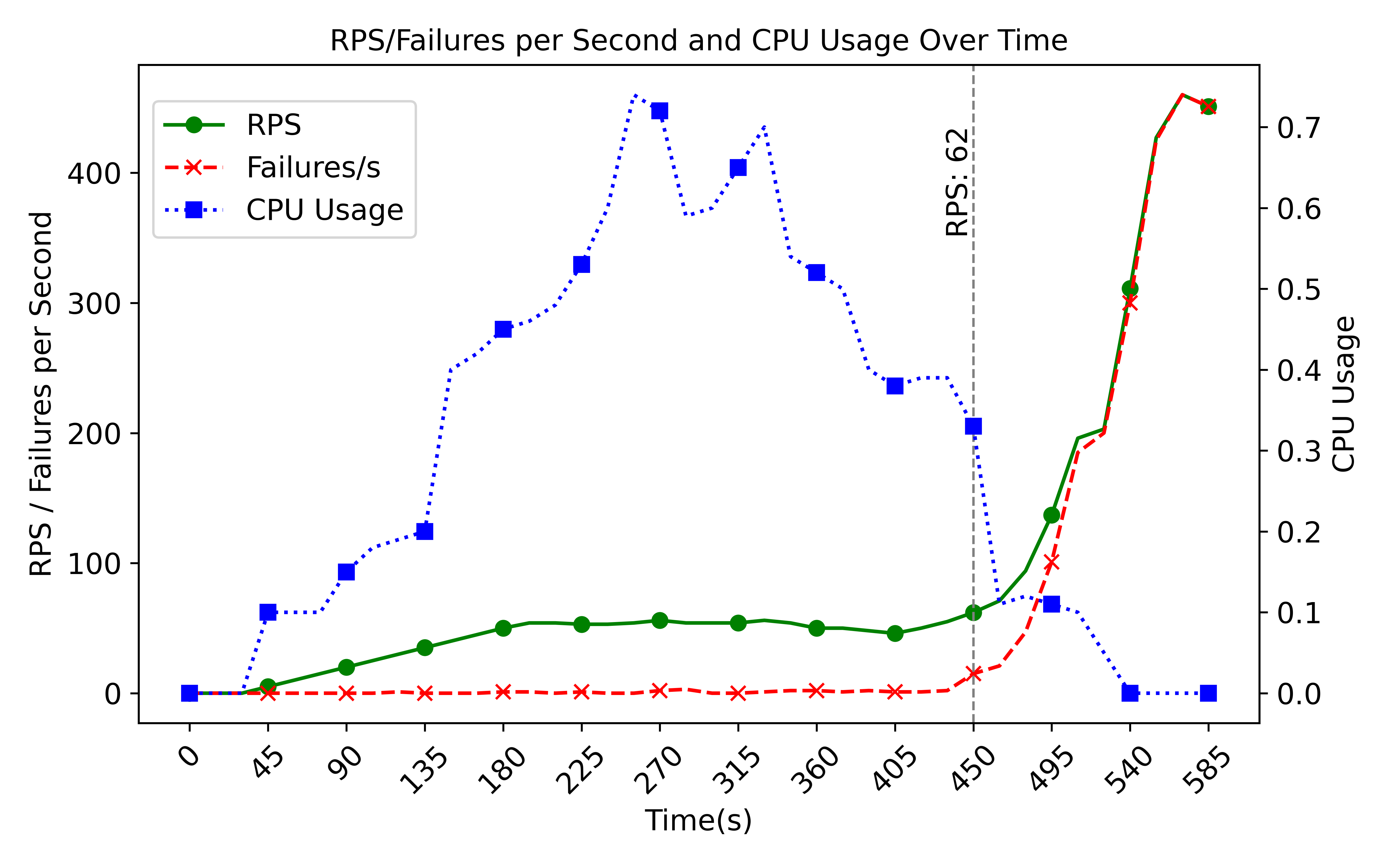}
    \caption{CPU Usage and Failure Rate}
    \label{fig: CPU Usage and Failure Rate}
\end{figure}

The analytical model is composed of two key components, namely the Metric Analyzer and the Pod Calculator. 
The \textbf{Metric Analyzer} takes the metrics collected by the load test, which include CPU usage per load and failure rates. The primary objective of the Metric Analyzer is to determine the maximum load(request per second) that a pod can handle without any performance degradation.

Figure \ref{fig: CPU Usage and Failure Rate} presents data on CPU utilization and the corresponding failure rate obtained through load testing of a prime number calculation application. Note that we have only deployed a single pod for this testing. As illustrated in Figure \ref{fig: CPU Usage and Failure Rate}, there is a notable decline in CPU usage beyond a certain point as illustrated in the vertical line, coinciding with an increase in failure rate. This inflection point marks the application's threshold for maximum load capacity that a pod can handle, beyond which it experiences crashes.

Subsequently, the \textbf{Pod Calculator} integrates data from the Metric Analyzer and the SLO to output the initial optimal number of pods required for deployment, ensuring alignment with the application's resource demands.

\[
\text{Initial Number of Pods} = \left\lceil \frac{\min(\text{RPS}_m)}{\max(\text{RPS}_{\text{Pod}_m})} \right\rceil
\]

Here's a breakdown of the components:

\begin{itemize}
    \item \(\min(\text{RPS}_m)\): Minimum RPS that the microservice \( m \) should handle. This is the SLO.
    \item \(\max(\text{RPS}_{\text{Pod}_m})\): Maximum RPS that a Pod belonging to microservice \( m \) can handle. This value is calculated using metric analyzer as depicted in figure \ref{fig: CPU Usage and Failure Rate}
    \item \(\left\lceil \cdot \right\rceil\): Ceiling function, which rounds up to the nearest integer.
\end{itemize}

\subsection{\textbf{Cost optimal Resource Allocation}}This paper addresses the critical challenge of optimizing monetary costs while maintaining user-defined performance for microservices applications. \medskip

\noindent\textbf{Optimization Engine:} The optimization Engine is responsible for addressing those challenges and it mainly consists of the following three components. 

\textbf{Predictor:} The predictor forecasts the spot price for each predefined node type unless specifically defined by the user. It analyzes time series data from the past three months, retrieved in real-time from the AWS Spot Pricing History API \cite{describe8:online}, and continuously updates its model to predict spot prices. Given the high variability across instance types \cite{baldominos2022aws}, Prophet \cite{prophet} is employed for its flexibility in modeling complex seasonality and its ability to adjust dynamically to real-time data. This method outperforms static, pre-trained models, offering more accurate short-term forecasts, especially in environments with rapid price fluctuations.

\textbf{Optimizer:} The optimizer leverages the predicted spot prices to identify the best node combination that corresponds to the given workload, utilizing a brute force, a greedy approach, or a genetic algorithm-based Pareto optimization. 

\begin{figure}[!h]
    \centering
    \includegraphics[scale=0.45]{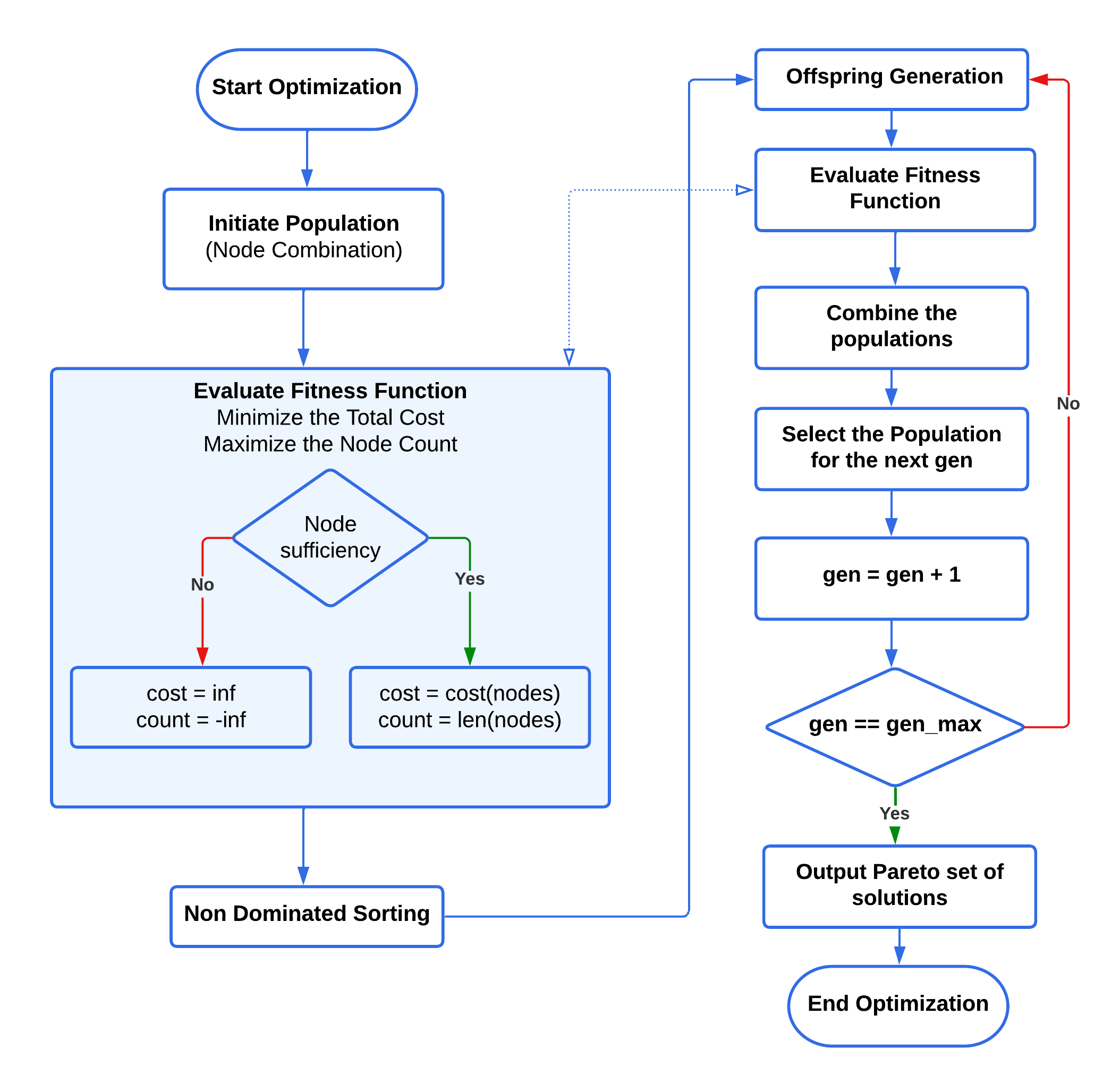}
    \caption{Pareto Optimization}
    \label{fig: Pareto Optimization}
\end{figure}

NSGA-II\cite{Afastand1:online} based pareto optimization solutions have demonstrated the best trade-off between cost and performance, which is one of the vital research outcomes of our study. 
The following diagram \ref{fig: Pareto Optimization} illustrates the flow of the optimization approach.

The fitness function identifies the optimal set of nodes that minimize costs and maximize the number of nodes, ensuring system availability through diverse node allocation across availability zones and instance families. It ensures node sufficiency by comparing each node combination's resource availability with the current workload's requirements. This topic will be further discussed in the discussion section.

\textbf{Cost Model:} The cost model calculates the cost of a selected set of nodes and returns the cost per unit time (1 hour) to the given pod count. \medskip

\(C(m) = \sum_i^{SerType}C_{m,i}^{Ser}S_m^i \)
\medskip

\begin{itemize}
  \item\(C_{m,i}^{Ser}\): Cloud server usage cost for server type $i$ per unit in time period $m$.
  \item\(S_m^i\): Amount of server usage of type $i$ in time period $m$. 
  \item\(SerType\): Set of all server types considered for deployment in the cloud environment.
\end{itemize} \medskip

\(C_{m,i}^{Ser}\) is calculated using the real-time spot price predictor model that we implemented.
\medskip

% \noindent\textbf{Node Allocator:} Based on the configuration provided by the optimization engine the node allocator requests spot instances from the spot market and assigns them to the private subnet in the virtual private cloud inside the public cloud which is provisioned by the provisioner.
% Once the spot instances are acquired, the node allocator proceeds to configure the master node and other worker nodes, these worker nodes are then assigned to the cluster where the microservices application is deployed.

According to the SpotKube approach, currently, the cluster is ready for deployment. Once the application is deployed, based on the incoming traffic cluster needs to be dynamically scaled which is explained in detail in the next section.

\subsection{\textbf{Dynamic Scaling of the Cluster}} Microservices applications require efficient management and automation for optimal performance, and container orchestration tools like Kubernetes are essential for handling them. SpotKube’s elastic scaler addresses the challenges of cost optimization in cluster auto-scaling and maintaining system availability, particularly with the abrupt termination of spot instances. 
\medskip
	
\noindent\textbf{Elastic Scaler:} Elastic Scaler operates by periodically notifying the optimization engine about the current configuration, enabling the cluster to adjust to the application's needs. The elastic scaler continuously communicates with the Kubernetes API server and compares any changes in the current configuration with the initial configuration. Upon detecting a valid change it will collect more data from the API server and apply certain heuristics to determine if the configuration change is significant. If the change is deemed significant, the optimization engine will be invoked with the current configuration.

Further, to detect sudden spot termination signals, the elastic scaler has implemented a service utilizing the AWS node termination handler \cite{AWS-node-termination-handler}. This ensures that when a sudden spot instance termination occurs, the relevant pods will be rescheduled gracefully, and the optimization engine will be notified of the unavailability of the specific instance type.

Based on the current pod count and spot termination details, the optimization engine will recommend a new optimal node combination and resources will then be attached to or detached from the cluster accordingly.

% Following flowchart \ref{fig: Elastic Scaler} illustrates the high-level workflow of the elastic scaler.

% \begin{figure}[!h]
%     \centering
%     \includegraphics[scale=0.45]{Images/ieee_cloud/elastic_scaler.png}
%     \caption{Elastic Scaler}
%     \label{fig: Elastic Scaler}
% \end{figure} 

% ------------------------- Work Done ------------------------------------------------
\section{EVALUATION}

\subsection{EXPERIMENTAL SETUP}

SpotKube can be extended to other cloud vendors and multi-cloud environments, but for demonstration purposes, we have integrated it with AWS services.

\subsubsection{\textbf{Hardware setup}}
\paragraph{Kubernetes Manager Node} The Kubernetes manager node was provisioned on an AWS EC2 instance of type t3.medium which aligns with the requirements of managing the Kubernetes cluster efficiently.

\paragraph{Kubernetes Worker Node} A t3.medium EC2 instance was designated as the Kubernetes worker node.

\paragraph{SpotKube Services Node}The SpotKube services, responsible for managing AWS spot instances within the Kubernetes cluster, were deployed on a t2.micro EC2 instance.

% \medskip
\subsubsection{\textbf{Microservice applications}}

The following four HTTP-based microservices are selected from popular open-source projects for the experiments.

\begin{itemize}
    \item Gateway service (Network heavy application) \cite{kantancoding_microservices_python}
    \item Auth service (Database access) \cite{kantancoding_microservices_python}
    \item Guest-book Go service (Web access) \cite{kubernetes_guestbook_go}
    \item Prime-app service (Compute intensive)
\end{itemize}

% \medskip
\subsubsection{\textbf{Load testing}}
The microservices used in this experiment operate asynchronously. For synchronous services, mock implementations of dependencies would be necessary before load testing, as suggested by Jindal et al. \cite{jindal2019performance}. To conduct load testing, SpotKube utilizes Locust, which linearly increases the number of users spawned per second, while Prometheus collects CPU metrics.

% \medskip
\subsubsection{\textbf{Deployment Strategy}}

Resource limits were imposed on each container for simplicity, and Minikube was selected as the K8s cluster for the experiment. Two deployment strategies were employed followed by load testing of the applications to evaluate the proposed approach:
\begin{itemize}
    \item Default - Horizontal Pod Autoscaler (HPA) and Cluster Autoscaler (CA) with AWS Autoscaling Group
    \item SpotKube - HPA and SpotKube
\end{itemize}

\subsection{EXPERIMENTAL EVALUATION}

\subsubsection{\textbf{Analytical Model Validation}}

For this evaluation, we set a user-defined Service Level Objective (SLO) for all applications, targeting a minimum support of 50 requests per second.

\begin{figure}[!h]
    \centering
    \includegraphics[scale=0.42]{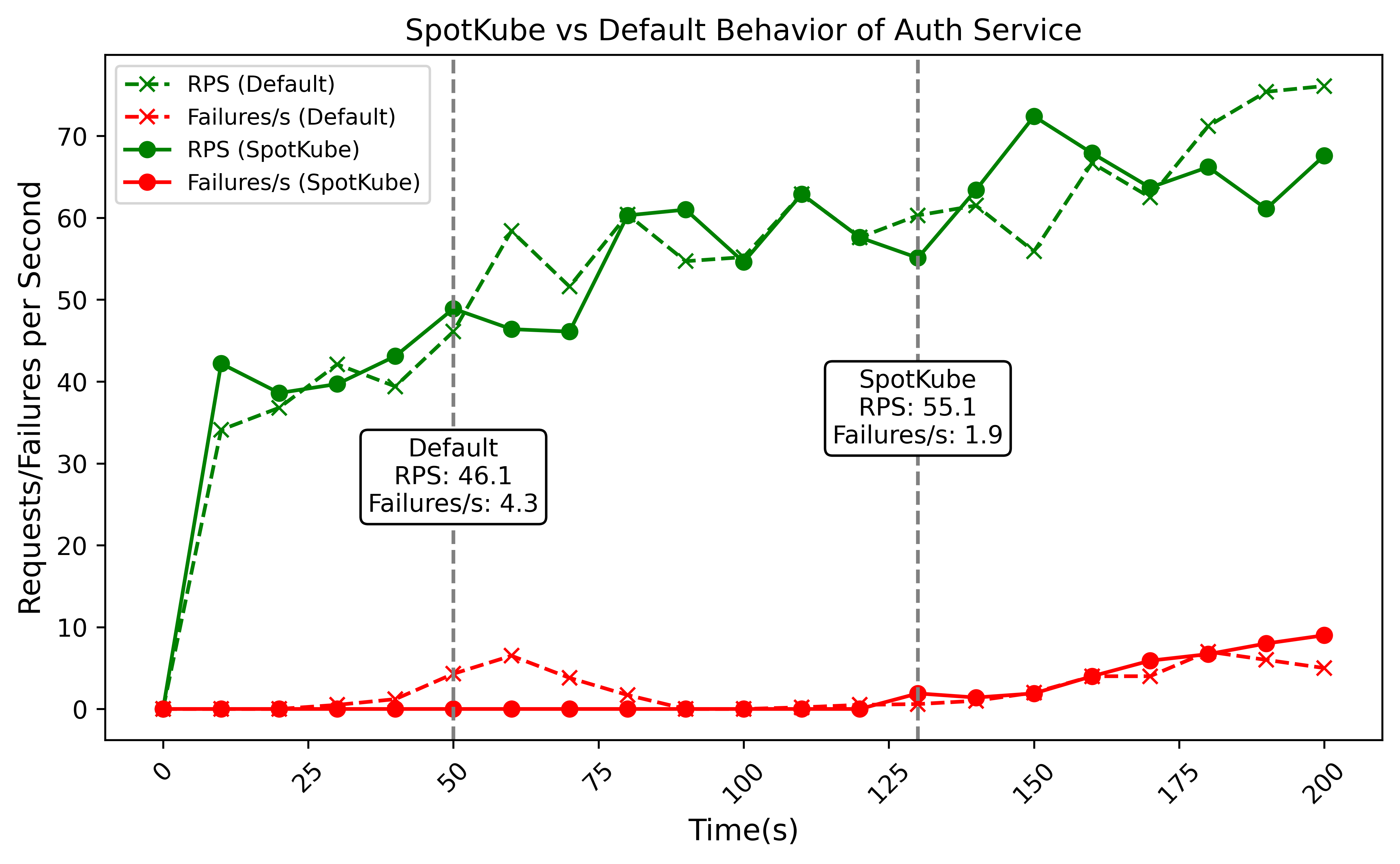}
    \caption{SpotKube vs Default Behavior of Auth Service}
    \label{fig: RPS and Failure Rate}
\end{figure}

Figure \ref{fig: RPS and Failure Rate} presents the load test results for the Auth service, which was deployed using both default and SpotKube strategies. The results indicate that the default deployment setup experienced some failures before meeting the specified SLO. In contrast, the SpotKube-based deployment successfully achieved the user-defined objectives.

Table \ref{tab:combined_behavior} compares two deployment strategies, highlighting the RPS at which failures occur with the Service Level Objective (SLO). The results indicate that deploying microservices using SpotKube is more effective than an ad-hoc deployment approach. \\

\begin{table}[!h]
\centering
\resizebox{\columnwidth}{!}{%
\begin{tabular}{|l|r|r|rr|rr|}
\hline
\multirow{2}{*}{Service} &
  \multirow{2}{*}{\begin{tabular}[c]{@{}c@{}}Spawn\\ Rate \end{tabular}} &
  \multirow{2}{*}{\begin{tabular}[c]{@{}c@{}}SLO\\ (RPS)\end{tabular}} &
  \multicolumn{2}{c|}{Default} &
  \multicolumn{2}{c|}{Spotkube} \\ \cline{4-7} 
 &  &  & 
  \begin{tabular}[c]{@{}c@{}}RPS\end{tabular} &
  \begin{tabular}[c]{@{}c@{}}Failures\end{tabular} &
  \begin{tabular}[c]{@{}c@{}}RPS\end{tabular} &
  \begin{tabular}[c]{@{}c@{}}Failures\end{tabular} \\ \hline
Auth &
  10 &
  50 &
  46.1 &
  4.3 &
  55.1 &
  1.9 \\ \hline
Gateway &
  10 &
  50 &
  46.5 &
  6.5 &
  54.1 &
  1.9 \\ \hline
Guestbook &
  10 &
  50 &
  43.3 &
  2.8 &
  54.1 &
  1.6 \\ \hline
Prime-app &
  2 &
  50 &
  51.2 &
  0 &
  62.5 &
  0 \\ \hline
\end{tabular}%
}
\caption{Analytical Model Validation}
\label{tab:combined_behavior}
\end{table}

% \begin{table}[!h]
% \centering
% \resizebox{\columnwidth}{!}{%
% \begin{tabular}{|l|rr|r|rr|rr|}
% \hline
% \multirow{2}{*}{Service} &
%   \multicolumn{2}{c|}{Load testing} &
%   \multirow{2}{*}{\begin{tabular}[c]{@{}c@{}}SLO\\ (RPS)\end{tabular}} &
%   \multicolumn{2}{c|}{Default} &
%   \multicolumn{2}{c|}{Spotkube} \\ \cline{2-3} \cline{5-8} 
%  &
%   \begin{tabular}[c]{@{}c@{}}Number\\ of users\end{tabular} &
%   \begin{tabular}[c]{@{}c@{}}Spawn\\ rate\end{tabular} &
%    &
%   \begin{tabular}[c]{@{}c@{}}RPS when\\ observe failures\end{tabular} &
%   \begin{tabular}[c]{@{}c@{}}Number of\\ failures\end{tabular} &
%   \begin{tabular}[c]{@{}c@{}}RPS when\\ observe failures\end{tabular} &
%   \begin{tabular}[c]{@{}c@{}}Number of\\ failures\end{tabular} \\ \hline
% Auth &
%   \multicolumn{1}{r|}{1000} &
%   10 &
%   45 &
%   \multicolumn{1}{r|}{43.2} &
%   1.6 &
%   \multicolumn{1}{r|}{49.4} &
%   1.5 \\ \hline
% Gateway &
%   \multicolumn{1}{r|}{1000} &
%   10 &
%   50 &
%   \multicolumn{1}{r|}{46.5} &
%   6.5 &
%   \multicolumn{1}{r|}{55.1} &
%   1.9 \\ \hline
% Guestbook &
%   \multicolumn{1}{r|}{1000} &
%   10 &
%   50 &
%   \multicolumn{1}{r|}{43.3} &
%   2.8 &
%   \multicolumn{1}{r|}{54.1} &
%   1.6 \\ \hline
% Prime-app &
%   \multicolumn{1}{r|}{500} &
%   2 &
%   60 &
%   \multicolumn{1}{r|}{51.2} &
%   0 &
%   \multicolumn{1}{r|}{72.5} &
%   0 \\ \hline
% \end{tabular}%
% }
% \caption{Analytical Model Validation}
% \label{tab:combined_behavior}
% \end{table}

\subsubsection{\textbf{Spot Predictor Validation}}

\begin{figure}[!h]
    \centering
    \includegraphics[scale=0.38]{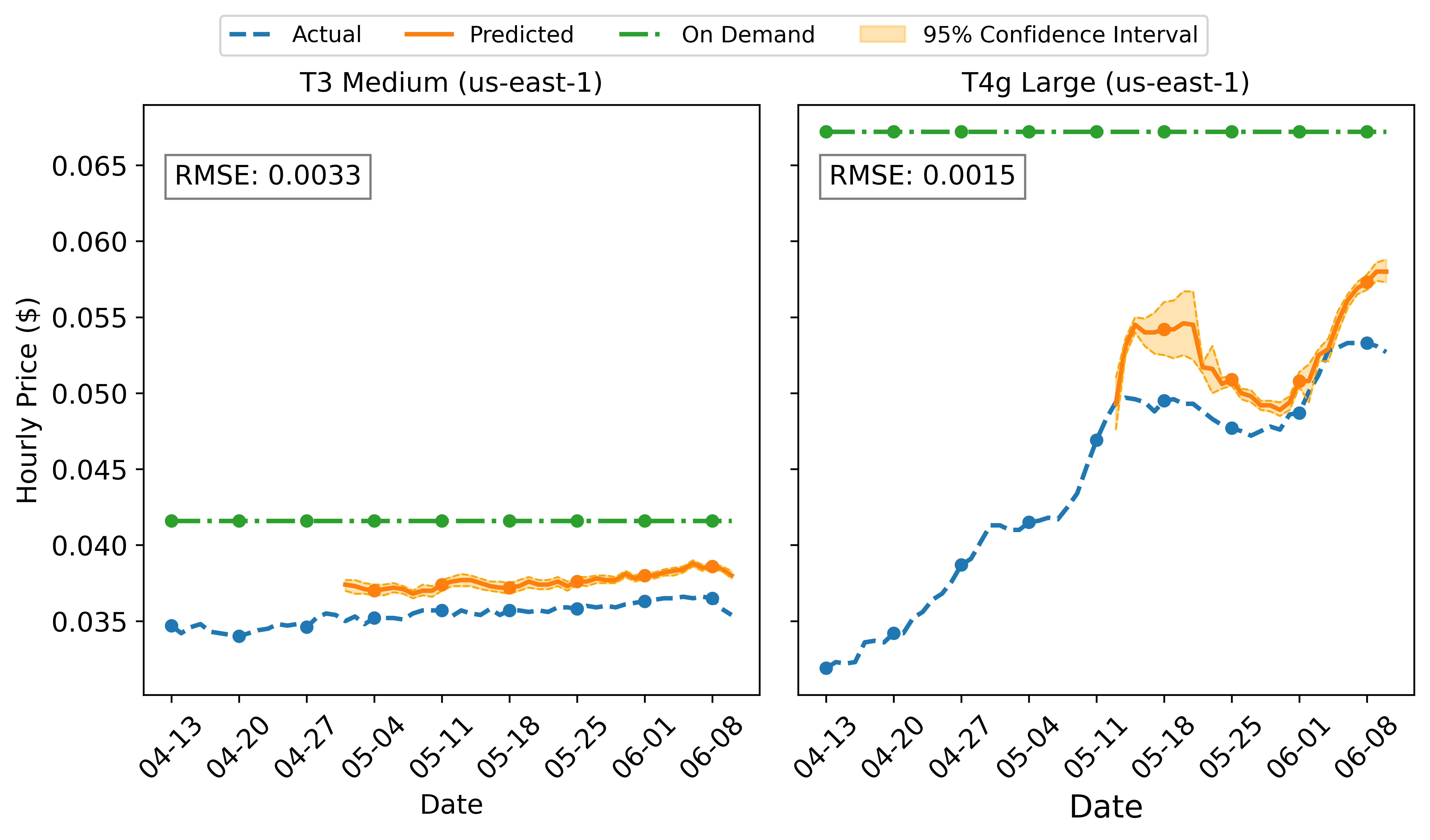}
    \caption{Spot Price Prediction}
    \label{fig: Spot Price Prediction}
\end{figure}

SpotKube relies on accurate spot price forecasts to request spot instances from the AWS spot market, helping to minimize unnecessary instance terminations. As shown in Figure \ref{fig: Spot Price Prediction}, we evaluated our Prophet-based spot price forecasting model on two instance types. The model’s low Root Mean Square Error (RMSE) values indicate high forecasting accuracy, supporting reliable spot instance bidding. Furthermore, the narrow 95\% confidence intervals reinforce the robustness of the approach, confirming their reliability for spot price prediction. 

% \medskip
\subsubsection{\textbf{Optimization Algorithms Comparison}}

\begin{figure*}[!h] 
    \centering
    \begin{minipage}[t]{0.32\textwidth} 
        \centering 
        \includegraphics[width=\textwidth]{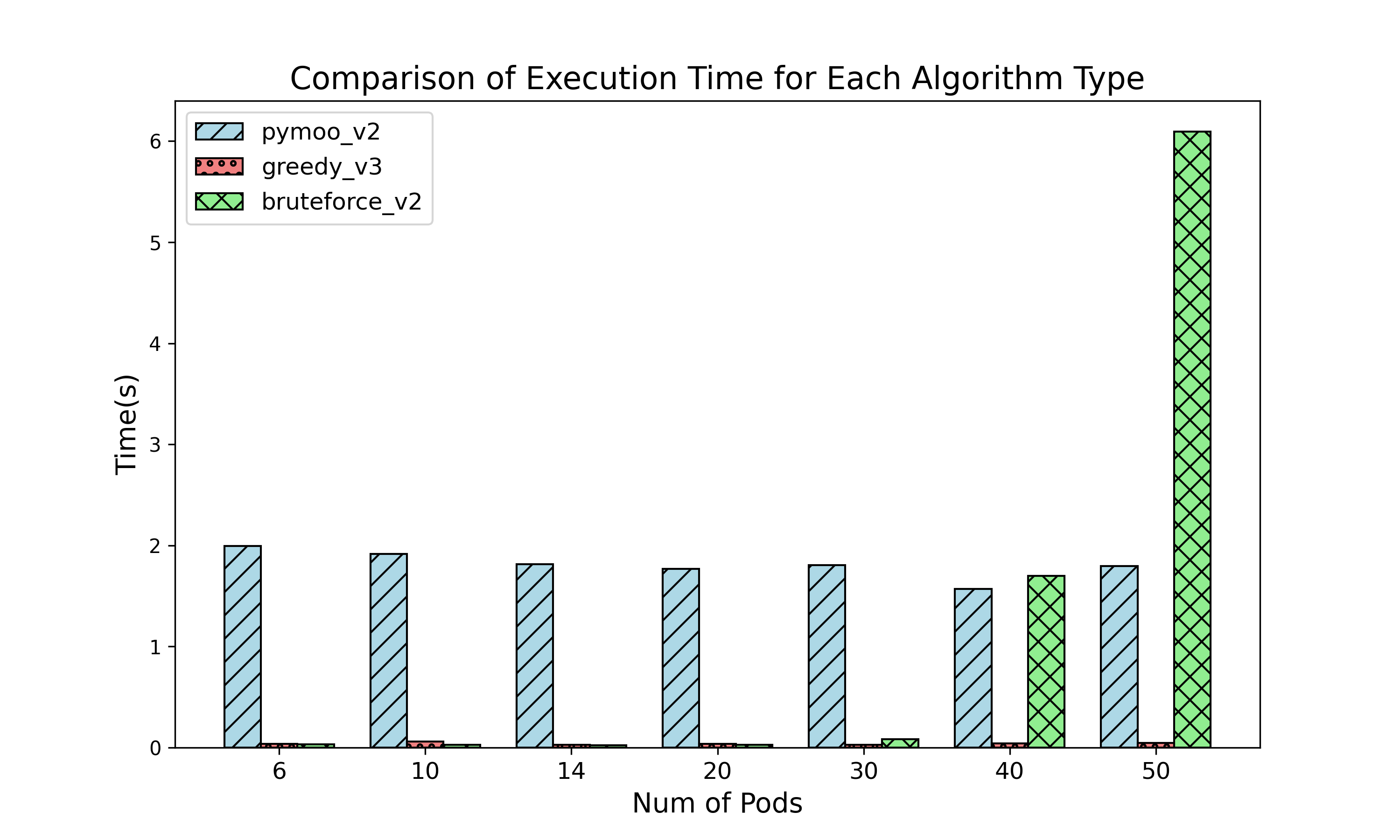} 
        \subcaption{Time(s)}
    \end{minipage}
    % \hspace{0.01\textwidth}
    \begin{minipage}[t]{0.32\textwidth} 
        \centering 
        \includegraphics[width=\textwidth]{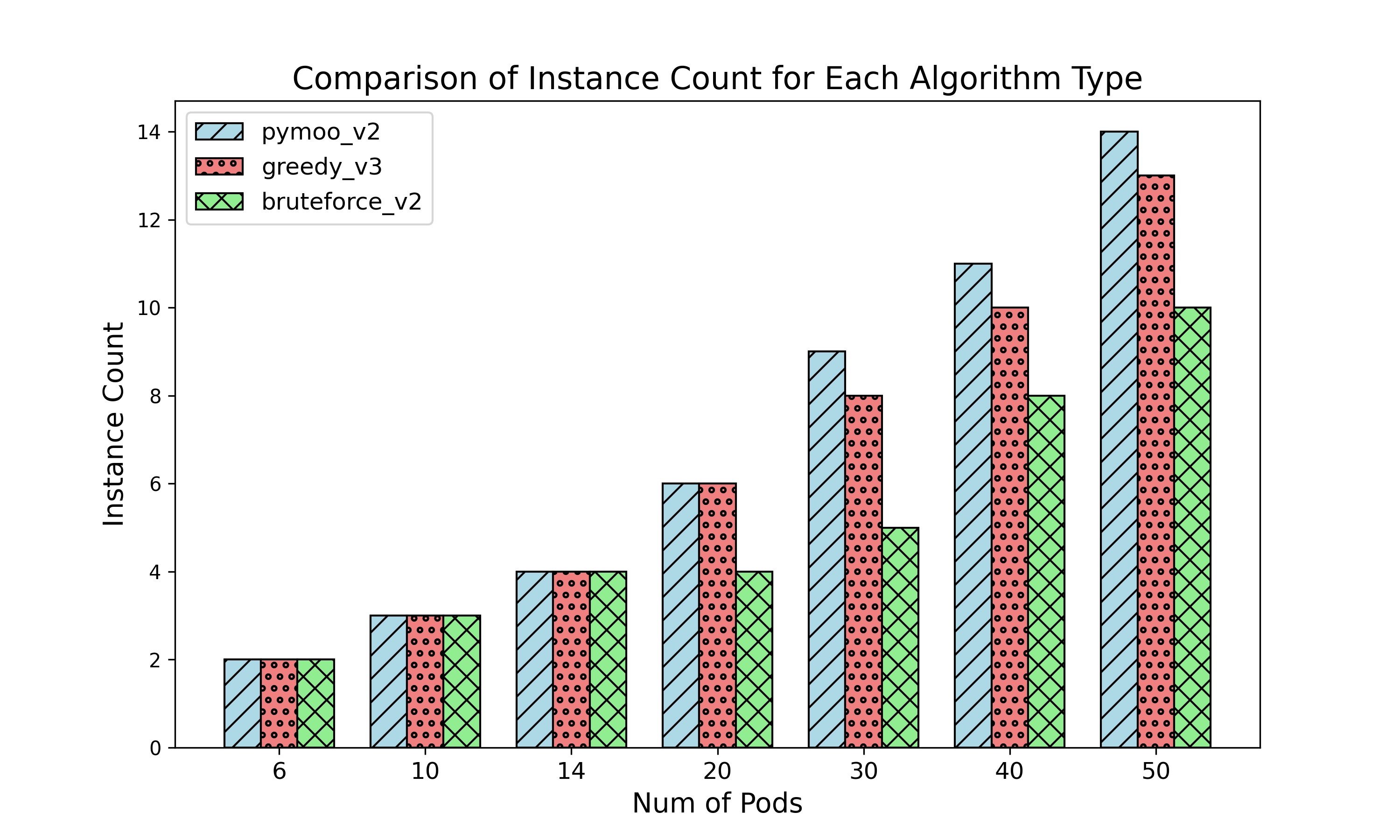} 
        \subcaption{Instance Count}
    \end{minipage} 
    % \hspace{0.01\textwidth} 
    \begin{minipage}[t]{0.32\textwidth} 
        \centering 
        \includegraphics[width=\textwidth]{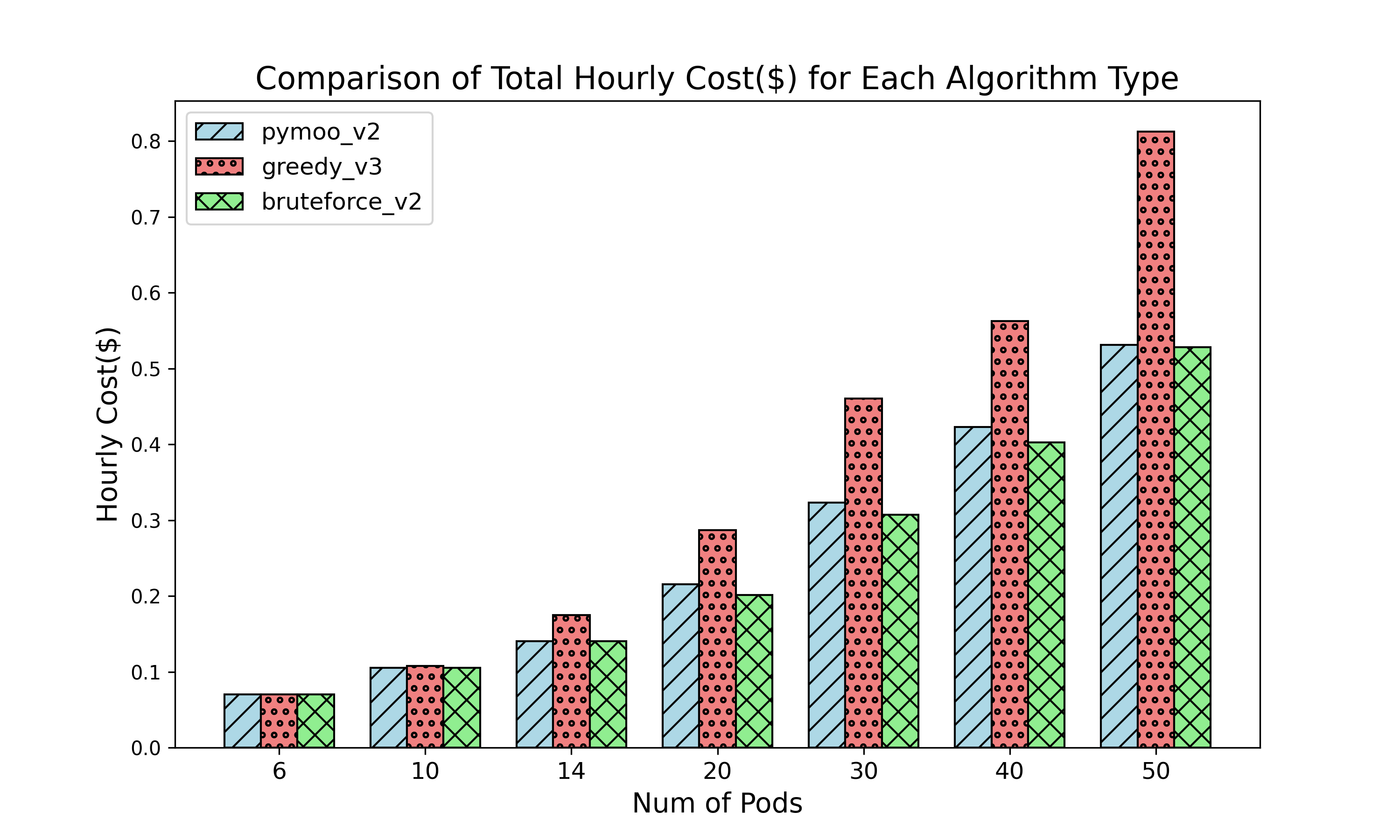} 
        \subcaption{Hourly Cost(\$)}
    \end{minipage}
    \caption{Optimization Algorithm Comparison}
    \label{fig: opt-algo-comparison}
\end{figure*}

We experimented to compare the performance of three optimization algorithms: Bruteforce, Greedy approach, and NSGA-II-based Pareto optimization. The evaluation focused on three criteria: execution time, total cost, and the number of nodes. The algorithms' performance was analyzed with varying numbers of pods to understand their efficiency and effectiveness relative to the problem's scale. \\
The primary objective was to identify the optimal trade-off between cost and performance by determining the best set of node combinations, where cost is calculated using a defined cost model, and performance is ensured by distributing pods across multiple nodes to maintain system availability even during sudden spot instance terminations. \\
According to Figure \ref{fig: opt-algo-comparison}, results indicated that Pareto optimization consistently produced the best set of nodes in a uniform time frame, regardless of the pod count, achieving minimal cost and maximal node distribution.

% \begin{comment}
% \begin{table}[H]
% \resizebox{\columnwidth}{!}{%
% % \begin{tabularx}{\textwidth}{|p{0.1\textwidth}|X|X|X|X|X|}
% \begin{tabular}{|c|c|c|c|c|c|}
% \hline
% \textbf{Service} & \textbf{Initial pod count} & \textbf{Final pod count} & \textbf{Initial node count} & \textbf{Initial node count} & \textbf{Time taken (mins)}\\
% \hline
% Prime app & 2 & 6 & 1 (t3.medium) & 2 (t3.medium, t3.medium) & 8.1 \\ \hline
% Prime app & 2 & 10 & 1 (t3.medium) & 3 (t3.medium, t3.medium, m6g.medium) & 7.2  \\ \hline
% Gateway & 8 & 4 & 2 (t3.medium, t3.medium) & 1 (t3.medium) & 7.7  \\ \hline
% Gateway & 8 & 15 & 2 (t3.medium, t3.medium) & 4 (t3.medium, t3.medium, c6a.large, t4g.large) & 8.3 \\ \hline
% \end{tabular}
% }
% \end{table}
% \end{comment}

\subsubsection{\textbf{Elastic Scaler Validation}}

\paragraph{\textbf{Validation of Cost-Effectiveness of the Elastic Scaler}}
To assess the cost-effectiveness of our Elastic Scaler and SpotKube solution, we compared it to an Amazon EKS cluster \cite{awsEKS} based on spot instances, with both the Cluster Autoscaler (CA) and Horizontal Pod Autoscaler (HPA) enabled. The experiment involved deploying two applications, the Prime App and the Gateway Service, with requests sent to both applications according to a predetermined pattern over 24 hours.

Resource limits for each pod were set to 0.5 vCPU and 1 GiB of memory, with the node group comprising t3.medium, c6a.large, t4g.large, and c6g.xlarge spot instance types. 
Post-experiment analysis revealed that the average pod count per hour was approximately 50 for both deployments. The average spot instance count and the associated costs were documented, with comparative costs for the Amazon EKS cluster and the SpotKube cluster presented in Table \ref{fig: cost-effectiveness}.

\begin{table}[H]
\resizebox{\columnwidth}{!}{%
\begin{tabular}{|cc|cc|}
\hline
\multicolumn{2}{|c|}{EKS (Spot Instance) \& CA}                                     & \multicolumn{2}{c|}{SpotKube}                                 \\ \hline
\multicolumn{1}{|c|}{Service}             & Cost (Hourly \$) & \multicolumn{1}{c|}{Service}              & Cost (Hourly \$) \\ \hline
\multicolumn{1}{|c|}{Amazon EKS}          & 0.10              & \multicolumn{1}{c|}{VPC}                  & 0                 \\ \hline
\multicolumn{1}{|c|}{Elastic IP x 1} & 0             & \multicolumn{1}{c|}{Elastic IP x 2}  & 0.01             \\ \hline
\multicolumn{1}{|c|}{}     &                   & \multicolumn{1}{c|}{EC2 (t3.medium) x 1}  & 0.0416             \\ \hline
\multicolumn{1}{|c|}{}      &                 & \multicolumn{1}{c|}{EC2 (t2.micro) x 1}   & 0.0116              \\ \hline
\multicolumn{2}{|c|}{}                                        & \multicolumn{1}{c|}{}           &             \\ \hline
\multicolumn{1}{|c|}{Spot (t3.medium) x 0}      & 0.00                 & \multicolumn{1}{c|}{Spot (t3.medium) x 8}   & 0.1328             \\ \hline
\multicolumn{1}{|c|}{Spot (c6a.large) x 0}      & 0.00                 & \multicolumn{1}{c|}{Spot (c6a.large) x 2}   & 0.0610              \\ \hline
\multicolumn{1}{|c|}{Spot (t4g.large) x 1}      & 0.026                 & \multicolumn{1}{c|}{Spot (t4g.large) x 1}   & 0.0268              \\ \hline
\multicolumn{1}{|c|}{Spot (c6g.xlarge) x 6}      & 0.326                 & \multicolumn{1}{c|}{Spot (c6g.xlarge) x 1}   & 0.0544              \\ \hline
% \multicolumn{2}{|c|}{\multirow{2}{*}{}}                       & \multicolumn{1}{c|}{}             \\ \cline{3-4} 
\multicolumn{1}{|c|}{Total(Hourly \$)}               &  0.4524           & \multicolumn{1}{c|}{Total(Hourly \$)}                     & 0.3382             \\ \hline
\multicolumn{2}{|c|}{}                                        & \multicolumn{1}{c|}{}           &             \\ \hline
\multicolumn{1}{|c|}{\textbf{Total(Monthly \$})}               &  \textbf{330.25}           & \multicolumn{1}{c|}{\textbf{Total(Monthly \$})}                     & \textbf{246.88 }             \\ \hline
\end{tabular}
}
\caption{Cost effectiveness of the Elastic Scaler}
\label{fig: cost-effectiveness}
\end{table}

The findings reveal a significant cost saving of up to 25\% through the utilization of our SpotKube approach when compared to Amazon EKS with spot instances and Cluster Autoscaler. This highlights the cost-efficiency and economic advantages offered by our solution.

\begin{table*}[t]
\resizebox{2\columnwidth}{!}{%
% \begin{tabularx}{\textwidth}{|p{0.1\textwidth}|X|X|X|X|X|}
\begin{tabular}{|c|c|c|c|c|c|}
\hline
\textbf{Service} & \textbf{Initial pod count} & \textbf{Final pod count} & \textbf{Initial node count} & \textbf{Final node count} & \textbf{Time taken (mins)}\\
\hline
Prime app & 2 & 6 & 1 (t3.medium) & 2 (t3.medium, t3.medium) & 8.1 \\ \hline
Prime app & 2 & 10 & 1 (t3.medium) & 3 (t3.medium, t3.medium, m6g.medium) & 7.2  \\ \hline
Gateway & 8 & 4 & 2 (t3.medium, t3.medium) & 1 (t3.medium) & 7.7  \\ \hline
Gateway & 8 & 15 & 2 (t3.medium, t3.medium) & 4 (t3.medium, t3.medium, c6a.large, t4g.large) & 8.3 \\ \hline
\end{tabular}%
}
\caption{Elastic Scaler Efficiency Validation}
\label{tab: Elastic Scaler Efficiency Validation}
\end{table*}

\paragraph{\textbf{Elastic Scaler Efficiency Validation}}

To validate the effectiveness and efficiency of the Elastic Scalar, we deployed several microservices and enabled the HPA. For each pod, we assigned a compute-intensive task and closely monitored the environment using Grafana. The HPA was configured to scale up when the CPU usage exceeded 80\%. 

Gradually increasing the CPU pressure, we observed that a few nodes surpassed the 80\% CPU usage threshold. At this point, the Elastic Scaler detected the change and triggered the optimization engine.

As depicted in Table \ref{tab: Elastic Scaler Efficiency Validation} Elastic Scaler and the Optimization Engine decision were taken within 30 seconds while the node allocation process took around 6-7 minutes.  In comparison, the AWS Cluster Autoscaler with AWS Autoscaling group took less than 30 seconds to make scaling decisions, while the cloud provider required 2 to 8 minutes to create compute resources \cite{kanthimathinathan2022reducing} \cite{aws-eks-best-practices-cluster-autoscaler}. Consequently, further research is necessary to decrease the cluster autoscaling time and enhance the performance of the node allocation process.

%---------------------------Discussion-------------------------------------------

\section{DISCUSSION}

In this discussion, we critically analyze the effectiveness of SpotKube, a framework designed to tackle challenges related to cloud cost optimization. It employs innovative approaches such as application characterization, elastic cluster autoscaling, and the integration of spot pricing models.

\subsection{Effectiveness of Application Characterization} 

Application characterization is responsible for determining the initial number of pods (replicas) that need to be deployed without violating the user-defined SLOs. Due to the dynamic nature of microservice applications, SpotKube is only considered for CPU-intensive applications, and for deployment, homogeneous pods are considered. 
As described in the methodology section, each microservice is subjected to a load test, and results were analyzed to determine the required pod count without violating SLOs. The data we collected through the load test consists of CPU usage and the failure rate per given load (RPS). Through experiments, we identified that by analyzing the failure rates and CPU usage data related to the load test, we can determine the maximum RPS value that can be handled by a pod. Then, by mapping this RPS value with the SLO (Minimum RPS value that a microservice should handle), we were able to determine the initial pod count that needed to be deployed. 
Experimental results show that the default behavior of Kubernetes with the HPA encountered some failures during pod scaling, while SpotKube is able to satisfy the user-defined SLOs without any failure. Although this can be thought of as an over-provisioning mechanism, due to the usage of spot instances, the cost has been reduced significantly, and the application performance is maintained without any failures.

\subsection{Elastic Cluster Autoscaling through Pareto Optimization}

The resource allocation mechanism within SpotKube, dynamically adjusts the deployment of microservice applications, taking into consideration the current spot market prices and the demand of the application. The elastic scaler employs the default Horizontal Pod Autoscaler (HPA) for the autoscaling of pods. This process is triggered once the total CPU utilization of the cluster surpasses an 80\% threshold or falls below a 30\% threshold, at which point the optimization engine is invoked to perform the necessary resource allocation or deallocation within the cluster. The determination of these specific threshold values, 80\% and 30\%, was informed by an extensive review of existing literature as well as iterative trials and errors, to maintain the required performance of the application. 

A significant contribution of SpotKube is the utilization of a genetic algorithm for Pareto optimization, which facilitates an exploration of the cost-performance trade-offs to identify optimal deployment configurations. The experimental evaluation of this approach highlights its efficacy in identifying efficient solutions, with a particular focus on the algorithm's execution time. According to the Time Comparison graph, Pareto optimization consistently determines the most cost-effective and performance-optimal spot instances that should be attached to the cluster within a nearly uniform duration across various deployment scenarios. This aspect of the optimization algorithm, which strategically determines the most cost-effective set of spot instances while maximizing the count of instances to distribute the workload broadly, represents a proactive strategy to mitigate the risk associated with the abrupt termination of spot instances.

The capability of the elastic scaler to dynamically scale clusters is evident, yet it is not without its limitations. Specifically, in the context of Amazon EKS, the autoscaler requires a minimum of 2 minutes to integrate a new node into the cluster, with the process potentially extending up to 8 minutes. Conversely, experimental observations have indicated that the Spotkube Elastic Scaler averages a scaling-up period of about 6-8 minutes. These experiments were conducted using an Ubuntu AMI, onto which all necessary software packages were installed via Ansible \cite{ansible}. This approach significantly leads to the extended duration required for scaling operations. We assume that employing a custom AMI, pre-equipped with all requisite packages, could notably accelerate the scaling process.

\subsection{Integration of Spot Pricing and Cost -Effectiveness of SpotKube}

SpotKube's most significant innovation lies in its strategic use of spot instances within cloud environments to optimize microservices deployment costs. Unlike traditional cost optimization approaches that may not fully leverage the cost-saving potential of spot pricing due to its inherent volatility, SpotKube integrates spot instance pricing models with Kubernetes. This enables dynamic cluster autoscaling that significantly reduces operational costs while maintaining service performance. A notable feature of SpotKube is its machine learning-based, real-time spot price prediction mechanism, which thoroughly analyzes the spot market for each instance type. The robustness of this predictive approach is evidenced by the low Root Mean Square Error (RMSE) values obtained from experimental validations.

Abrupt termination of spot instances is a significant challenge that can cause substantial downtime in the cluster, directly impacting system availability. To mitigate this, SpotKube employs a mix of novel and traditional techniques. One strategy is to strategically request spot instances at a bid higher than the forecasted spot price, yet still below the cost of on-demand instances. Another technique involves allocating a mix of spot instances from different instance families and availability zones. Furthermore, our optimization strategy focuses on minimizing costs while maximizing the total node count. By maximizing node count, SpotKube distributes the workload as much as possible, thereby maintaining system availability even with the sudden termination of a node. Additionally, SpotKube utilizes AWS's abrupt termination signals to handle node termination and pod rescheduling gracefully.

The findings illustrate SpotKube's efficiency in achieving cost reductions of over 25\% compared to other spot-instance-based deployment methodologies, specifically within the context of Amazon Elastic Kubernetes Service (EKS).
% ------------------------- Conclusion ------------------------------------------------
\section{CONCLUSION and FUTURE WORKS}
In this paper, we present SpotKube, an open-source Kubernetes-managed service that optimizes the deployment cost of microservices applications in public cloud environments. SpotKube facilitates the deployment of microservices by analyzing application characteristics to maximize cost savings through cluster autoscaling and transient pricing models like AWS spot pricing. It also effectively handles abrupt spot instance terminations to maintain high availability. Our experimental results demonstrate that SpotKube can achieve cost savings of up to 25\% compared to other spot-instance-based approaches while still meeting user-defined SLOs.

As future work, we plan to enhance SpotKube's capabilities to support heterogeneous pod deployment. This will enable SpotKube to optimize the deployment of applications that use different hardware configurations, enabling better resource utilization and cost savings. We aim to enhance the performance of Elastic Scalar by minimizing the time it takes for scaling operations.

In summary, this research demonstrates the potential of leveraging spot pricing in public cloud environments to reduce costs through cluster auto-scaling while maintaining application performance, and future works aim to improve and expand SpotKube's capabilities to further optimize microservices deployment in a cost-optimized manner.

\addtolength{\textheight}{-12cm}   % This command serves to balance the column lengths

%%%%%%%%%%%%%%%%%%%%%%%%%%%%%%%%%%%%%%%%%%%%%%%%%%%%%%%%%%%%%%%%%%%%%%%%%%%%%%%%

%%%%%%%%%%%%%%%%%%%%%%%%%%%%%%%%%%%%%%%%%%%%%%%%%%%%%%%%%%%%%%%%%%%%%%%%%%%%%%%%


% \section*{ACKNOWLEDGMENT}

%%%%%%%%%%%%%%%%%%%%%%%%%%%%%%%%%%%%%%%%%%%%%%%%%%%%%%%%%%%%%%%%%%%%%%%%%%%%%%%%

\bibliographystyle{ieeetr}
\bibliography{bib}

\begin{thebibliography}{10}

\bibitem{TheFutur71:online}
SoftClouds, ``The future of software development with microservices.'' https://softclouds.medium.com/the-future-of-software-development-with-microservices-5f4d263272b0, Feb 2020.

\bibitem{jindal2019performance}
A.~Jindal, V.~Podolskiy, and M.~Gerndt, ``Performance modeling for cloud microservice applications,'' in {\em Proceedings of the 2019 ACM/SPEC International Conference on Performance Engineering}, pp.~25--32, 2019.

\bibitem{gu2021scheduling}
H.~Gu, X.~Li, M.~Liu, and S.~Wang, ``Scheduling method with adaptive learning for microservice workflows with hybrid resource provisioning,'' {\em International Journal of Machine Learning and Cybernetics}, vol.~12, no.~10, pp.~3037--3048, 2021.

\bibitem{Bento_Araujo_Barbosa_2023}
A.~Bento, F.~Araujo, and R.~Barbosa, ``Cost-availability aware scaling: Towards optimal scaling of cloud services - journal of grid computing,'' Dec 2023.

\bibitem{han2020refining}
J.~Han, Y.~Hong, and J.~Kim, ``Refining microservices placement employing workload profiling over multiple kubernetes clusters,'' {\em IEEE access}, vol.~8, pp.~192543--192556, 2020.

\bibitem{altmann2014cost}
J.~Altmann and M.~M. Kashef, ``Cost model based service placement in federated hybrid clouds,'' {\em Future Generation Computer Systems}, vol.~41, pp.~79--90, 2014.

\bibitem{baldominos2022aws}
A.~Baldominos~G{\'o}mez, Y.~Saez, D.~Quintana, and P.~Isasi, ``Aws predspot: Machine learning for predicting the price of spot instances in aws cloud,'' 2022.

\bibitem{Kubernet50:online}
``Kubernetes.'' \url{https://kubernetes.io/}.
\newblock (Accessed on 11/16/2022).

\bibitem{sami2020fscaler}
H.~Sami, A.~Mourad, H.~Otrok, and J.~Bentahar, ``Fscaler: Automatic resource scaling of containers in fog clusters using reinforcement learning,'' in {\em 2020 international wireless communications and mobile computing (IWCMC)}, pp.~1824--1829, IEEE, 2020.

\bibitem{rodriguez2020container}
M.~Rodriguez and R.~Buyya, ``Container orchestration with cost-efficient autoscaling in cloud computing environments,'' in {\em Handbook of research on multimedia cyber security}, pp.~190--213, IGI global, 2020.

\bibitem{yan2021hansel}
M.~Yan, X.~Liang, Z.~Lu, J.~Wu, and W.~Zhang, ``Hansel: Adaptive horizontal scaling of microservices using bi-lstm,'' {\em Applied Soft Computing}, vol.~105, p.~107216, 2021.

\bibitem{thurgood2019cloud}
B.~Thurgood and R.~G. Lennon, ``Cloud computing with kubernetes cluster elastic scaling,'' in {\em Proceedings of the 3rd International Conference on Future Networks and Distributed Systems}, pp.~1--7, 2019.

\bibitem{domanal2018efficient}
S.~G. Domanal and G.~R.~M. Reddy, ``An efficient cost optimized scheduling for spot instances in heterogeneous cloud environment,'' {\em Future Generation Computer Systems}, vol.~84, pp.~11--21, 2018.

\bibitem{LocustAm36:online}
``Locust - a modern load testing framework.'' \url{https://locust.io/}.
\newblock (Accessed on 04/05/2023).

\bibitem{Promethe23:online}
``Prometheus - monitoring system \& time series database.'' \url{https://prometheus.io/}.
\newblock (Accessed on 04/05/2023).

\bibitem{describe8:online}
``describe-spot-price-history — aws cli 1.27.116 command reference.'' \url{https://docs.aws.amazon.com/cli/latest/reference/ec2/describe-spot-price-history.html}.
\newblock (Accessed on 04/21/2023).

\bibitem{prophet}
{Facebook, Inc.}, ``{Prophet: Forecasting at Scale}.'' \url{https://facebook.github.io/prophet/}.
\newblock [Accessed: Jan. 20, 2024].

\bibitem{Afastand1:online}
``A fast and elitist multiobjective genetic algorithm: Nsga-ii | ieee journals \& magazine | ieee xplore.'' \url{https://ieeexplore.ieee.org/document/996017}.
\newblock (Accessed on 05/16/2023).

\bibitem{AWS-node-termination-handler}
{Amazon Web Services}, ``{AWS/AWS-node-termination-handler: Gracefully handle EC2 instance shutdown within kubernetes}.'' \url{https://github.com/aws/aws-node-termination-handler}.
\newblock Accessed: Feb. 19, 2024.

\bibitem{kantancoding_microservices_python}
KantanCoding, ``Microservices-python,'' 2024.
\newblock Accessed: 2024-04-11.

\bibitem{kubernetes_guestbook_go}
Kubernetes, ``Guestbook-go example on github,'' 2024.
\newblock Accessed: 2024-05-30.

\bibitem{awsEKS}
``{Amazon EKS - Managed Kubernetes Service}.'' Available: \url{https://aws.amazon.com/eks/}.
\newblock Accessed: Jul. 29, 2024.

\bibitem{kanthimathinathan2022reducing}
B.~R. Kanthimathinathan, {\em Reducing instance acquisition lag to improve scaling out in the kubernetes cluster}.
\newblock PhD thesis, Dublin, National College of Ireland, 2022.

\bibitem{aws-eks-best-practices-cluster-autoscaler}
{Amazon Web Services}, ``Cluster autoscaler - eks best practices guides.'' \url{https://aws.github.io/aws-eks-best-practices/cluster-autoscaling/}.

\bibitem{ansible}
``{Ansible Automation Platform}.'' Available: \url{https://www.ansible.com/}.
\newblock Accessed: Jul. 12, 2024.

\end{thebibliography}

\end{document}